%% file: main.tex
\PassOptionsToPackage{dvipsnames}{xcolor}
\documentclass[sigconf, nonacm]{acmart}

\newcommand\vldbdoi{XX.XX/XXX.XX}

\newcommand\vldbavailabilityurl{}

\usepackage[T1]{fontenc}

\usepackage{amssymb}
\usepackage{stmaryrd}
\usepackage{booktabs}
\usepackage{pifont}
\usepackage{xspace}
\usepackage{enumitem}
\usepackage[normalem]{ulem}
\usepackage{algorithm}
\usepackage{algpseudocode}

\usepackage{tabularx}
\usepackage{booktabs}

\newcommand{\cypherTech}{CypherGuard\xspace}

\newcommand{\wEAcb}{88.50}
\newcommand{\zsEAcb}{63.42}
\newcommand{\dEAcb}{25.08}
\newcommand{\execCB}{99.83}
\newcommand{\psjsCB}{0.933}
\newcommand{\zsExecCB}{93.52}%
\newcommand{\zsPsjsCB}{0.775}%
\newcommand{\sonnetZsEAcb}{71.76}
\newcommand{\dSonnetCB}{16.74}
\newcommand{\sonnetEAcb}{94.68}
\newcommand{\dSonnetLift}{22.92}
\newcommand{\dRL}{19}
\newcommand{\dCS}{14}
\newcommand{\dBest}{27}

\newcommand{\nQ}{2{,}348}

\theoremstyle{acmdefinition}
\newtheorem{example}{Example}[section]
\theoremstyle{acmplain}

\newenvironment{exampleO}%
  {\begin{example}}%
  {\hfill$\square$\end{example}}

\definecolor{promptframe}{gray}{0.55}
\definecolor{promptbg}{gray}{0.975}
\definecolor{prompttitle}{gray}{0.25}
\newsavebox{\cgpromptbox}
\newenvironment{promptbox}[1]{%
  \setlength{\fboxrule}{0.7pt}\setlength{\fboxsep}{7pt}%
  \begin{lrbox}{\cgpromptbox}%
  \begin{minipage}{\dimexpr\columnwidth-2\fboxsep-2\fboxrule\relax}%
    \small
    {\sffamily\bfseries\color{prompttitle}#1}\par
    \vspace{2pt}{\color{promptframe}\rule{\linewidth}{0.4pt}}\par\vspace{4pt}%
}{%
  \end{minipage}%
  \end{lrbox}%
  \fcolorbox{promptframe}{promptbg}{\usebox{\cgpromptbox}}%
}

\begin{document}

\title[CypherGuard]{CypherGuard: Database-Grounded Correction for
Reliable Text-to-Cypher}

\author{Zebin Chen}
\affiliation{%
  \institution{University of New South Wales}
  \city{Sydney}
  \country{AU}
}
\email{zebin.chen1@unsw.edu.au}

\author{Dong Wen}
\affiliation{%
  \institution{University of New South Wales}
  \city{Sydney}
  \country{AU}
}
\email{dong.wen@unsw.edu.au}

\author{Makbule Gulcin Ozsoy}
\affiliation{%
  \institution{Neo4j}
  \city{London}
  \country{UK}
}
\email{makbule.ozsoy@neo4j.com}

\author{Wenjie Zhang}
\affiliation{%
  \institution{University of New South Wales}
  \city{Sydney}
  \country{AU}
}
\email{wenjie.zhang@unsw.edu.au}

\newcommand{\rr}[1]{\textcolor{red}{#1}}
\newcommand{\rrr}[1]{\textcolor{blue}{#1}}
\newcommand{\del}[1]{\sout{#1}}% proposed deletion
\newcommand{\new}[1]{\textcolor{black}{#1}}% proposed addition

\begin{abstract}
Text-to-Cypher translates natural-language questions into Cypher queries over property graphs.
Graph database systems increasingly adopt this translation as a native query interface.
Large language models generate fluent Cypher but stay unreliable. 
On CypherBench a small model answers only $31.4\%$ of questions correctly, and even a strong model plateaus at $\sonnetZsEAcb\%$.
The dominant failure modes are structural rather than linguistic.
For example, a generated query may parse correctly and follow the schema, yet still return zero rows because it traverses a relationship against the stored graph direction.
To address this issue, we present \cypherTech, a training-free framework that detects and corrects these structural faults.
Our approach uses two core components: a graph-native error taxonomy to guide repair, and a mined rule library to guide query generation.
At inference time, \cypherTech first rewrites schema-detectable faults, then re-executes each draft and repairs it only when the query errors or returns nothing.
On all \nQ{} CypherBench test questions, across seven graphs, \cypherTech reaches $\wEAcb$\% overall execution accuracy, $+\dEAcb\%$ over the same-backbone zero-shot floor, surpassing an RL-fine-tuned 7B specialist and the strongest training-based system that we compare against.
\end{abstract}

\maketitle

% \pagestyle{\vldbpagestyle}
% \begingroup\small\noindent\raggedright\textbf{PVLDB Reference Format:}\\
% \vldbauthors. \vldbtitle. PVLDB, \vldbvolume(\vldbissue): \vldbpages, \vldbyear.\\
% \href{https://doi.org/\vldbdoi}{doi:\vldbdoi}
% \endgroup
% \begingroup
% \renewcommand\thefootnote{}\footnote{\noindent
% This work is licensed under the Creative Commons BY-NC-ND 4.0 International License. Visit \url{https://creativecommons.org/licenses/by-nc-nd/4.0/} to view a copy of this license. For any use beyond those covered by this license, obtain permission by emailing \href{mailto:info@vldb.org}{info@vldb.org}. Copyright is held by the owner/author(s). Publication rights licensed to the VLDB Endowment. \\
% \raggedright Proceedings of the VLDB Endowment, Vol. \vldbvolume, No. \vldbissue\ %
% ISSN 2150-8097. \\
% \href{https://doi.org/\vldbdoi}{doi:\vldbdoi} \\
% }\addtocounter{footnote}{-1}\endgroup

\ifdefempty{\vldbavailabilityurl}{}{
\vspace{.3cm}
\begingroup\small\noindent\raggedright\textbf{PVLDB Artifact Availability:}\\
The source code, data, and/or other artifacts have been made available at \url{\vldbavailabilityurl}.
\endgroup
}

\input{sections/01_introduction}
\input{sections/02_background}

\input{sections/03_method}

\input{sections/04_setup}
\input{sections/05_results}
\input{sections/06_ablations}
\input{sections/08_related}
\input{sections/09_conclusion}

\bibliographystyle{ACM-Reference-Format}
\bibliography{references}

\end{document}

%% file: sections/01_introduction.tex
\section{Introduction}\label{sec:intro}

Querying data in natural language is a long-standing ambition of the database community, pursued on the relational database~\cite{li2014nalir,saha2016athena,weir2020dbpal} and, as question answering over knowledge graphs, on the graph~\cite{berant2013freebase,unger2012templateqa}.
Large language models have made the drafting step cheap, and the relational side has industrialized accordingly. Text-to-SQL systems, open models, and benchmark-scale evaluations are standing subjects at data-management venues~\cite{gao2024dailsql,li2024codes,li2024nl2sql360}.
The graph side has reached the same moment. Graph DBMS vendors ship text-to-Cypher datasets, fine-tuned baselines, and integration tooling~\cite{ozsoy2025text2cypher,langchain2026cyphercorrector}, and GraphRAG pipelines demonstrate graph-backed retrieval at scale~\cite{edge2024graphrag}.

Text-to-Cypher methods are not yet reliable, and the techniques that made text-to-SQL trustworthy do not transfer to text-to-Cypher directly. Text-to-Cypher still falls far short of high execution accuracy. 
On Spider, the standard text-to-SQL benchmark, GPT-4 with retrieved few-shot examples reaches $86.6\%$ execution accuracy~\cite{gao2024dailsql}.
On CypherBench, an execution-scored text-to-Cypher benchmark, \texttt{claude-3.5-sonnet} reaches only $61.6\%$ and \texttt{gpt-4o} only $60.2\%$ zero-shot~\cite{feng2025cypherbench}.
RL-tuned Cypher-RI reaches only $69.6\%$~\cite{su2025cypherri} and CypherSmith reaches only $74.8\%$~\cite{zhang2026cyphersmith}.

\begin{figure}[t]
\centering
\small
\fbox{\parbox{0.95\linewidth}{
\textit{Q: How many countries does Germany maintain diplomatic
relations with?}\\[2pt]
\textcolor{red}{\ding{55}}~\textbf{Naive draft} (reads a mutual relation as
undirected)\\[1pt]
{\footnotesize\ttfamily MATCH (:Country \{name:'Germany'\}) \textcolor{red}{-}[:hasDiplomaticRelationWith] \textcolor{red}{-}
(c:Country) RETURN count(c)}\\[1pt]
{\footnotesize$\rightarrow$ \textcolor{red}{wrong}: stored both ways. The undirected match therefore counts each partner twice. Execution returns rows, no alarm.}\\[3pt]
\textcolor{ForestGreen}{\ding{51}}~\textbf{\cypherTech} (storage-reciprocity gate: the probe measures the relation reciprocal, and undirected becomes directed)\\[1pt]
{\footnotesize\ttfamily MATCH (:Country \{name:'Germany'\}) - [:hasDiplomaticRelationWith]\textcolor{ForestGreen}{->} (c:Country) RETURN count(c)}\\[1pt]
{\footnotesize$\rightarrow$ \textcolor{ForestGreen}{correct}.}
}}
\caption{A schema-consistent draft on a CypherBench relation runs without error
yet returns a wrong count, a fault no execution signal exposes.}
\label{fig:motivating}
\end{figure}

\autoref{fig:motivating} illustrates one common failure pattern in text2cypher.
Consider asking a model to count how many diplomatic partners a country has.
While \texttt{hasDiplomaticRelationWith} appears naturally mutual and undirected, the graph stores it as two directed edges.
Consequently, an undirected query pattern matches each relationship twice.
The generated query parses correctly, follows the schema, and executes without errors, yet silently returns doubled counts.
Prompting the model again cannot expose the double counting.
Similarly, simple execution fails to catch the bug because the query still returns rows.
Detecting this mismatch requires measuring the actual relationship storage pattern in the database.

% transition
% why still low accuracy
% the reason for the above examples is ... / the purpose of this paragraph - expection 
Prior work covers only some faults.
Models frequently make subtle errors in NL2Cypher task. They traverse relationships backwards, connect labels unsupported by the schema, inflate counts through path fan-out, or add unprompted ORDER BY and LIMIT clauses. However, syntactically and at the schema level, these queries remain valid.
Static schema checks accept such queries, and running them often raises no error either.
Prior work already checks a few of these faults.
LangChain's Cypher corrector rewrites relationship directions that contradict the schema~\cite{langchain2026cyphercorrector}.
\cite{ozsoy2026grammarschema} filter sampled candidates with an openCypher grammar and a relationship-direction check.
They also report an iterative refinement loop whose verifiers include a direction check, an execution check, and an LLM judge~\cite{ozsoy2025iterative}.
Existing taxonomies categorize these errors purely post-hoc~\cite{feng2025cypherbench,su2026gqlbench}.
However, no text-to-Cypher method uses these categories to steer query generation.

Existing work improves text-to-Cypher accuracy along two routes.
The first route trains the model on Cypher data.
\cite{su2025cypherri} fine-tunes the model with execution-based rewards using Reinforcement learning (RL).
Cypher pairs~\cite{zhang2026cyphersmith} trains the model on question via Supervised fine-tuning (SFT).
The second route leaves the model unchanged and works at inference time.
Multi-agent frameworks assign query generation, result checking, and repair to separate LLM agents~\cite{gusarov2025multiagent}.
Execution-feedback methods regenerate a query after the database reports an error~\cite{jung2026ras}.
Both routes have limitations.
Training needs gold-labeled data.
RL requires a gold query and an executable database for every training question, and has been trained on a benchmark training split.
SFT faces a data shortage. Public Cypher datasets hold at most tens of thousands of examples, and recent work synthesizes hundreds of thousands with LLMs. % SFT -- make it more clear
Inference-time methods take their correction signal from the database's error message or from an LLM critic's judgment of the returned rows.
A silent fault raises no error, and the critic's judgment of the returned rows is not checked against the database.

We present \cypherTech, an admission-control layer between the LLM and the graph database. Admission here means screening queries and rewrites. 
Our pipeline consists of three stages:
(1) \emph{Draft}: The LLM generates an initial query guided by mined error-prevention rules and retrieved demonstration examples.
(2) \emph{Schema Fixes}: A deterministic module corrects structural faults revealed by schema constraints or storage statistics, such as an undeclared relationship type. 
(3) \emph{Execution Loop}: If a query fails or returns an empty result, the LLM identifies the error type from our Cypher taxonomy and rewrites the query.
% %
% A backwards graph traversal matches nothing. One of the most reported Cypher error classes~\cite{feng2025cypherbench} therefore announces itself as exactly the empty result that our Execution Loop watches for.
The layer's rule library is mined once per dataset, by prompting an LLM agent over the backbone's failures on that dataset's train split. 
% Moving the layer to a new graph or backbone therefore needs no retraining, and the layer's per-question call budget is fixed.

\noindent Our contributions are:
\begin{itemize}[leftmargin=*, topsep=2pt, itemsep=2pt, parsep=0pt]
  \item \textbf{Repairs backed by database signals.}
  We frame reliable text-to-Cypher as admission control.
  The framework consults multiple signals: question phrasing, schema constraints, stored-data statistics, and execution feedback.
  Each signal triggers only targeted repairs for the specific error it detects.
  Furthermore, the system accepts a proposed rewrite only if it passes re-verification or replaces an already-failed query.

  \item \textbf{A transferable Cypher error taxonomy.}
  We distil recurring text-to-Cypher failures into eight families and $32$ fine-grained codes. Seven families, spanning $28$ codes, name faults that leave the query executable, such as a reversed relationship direction, a wrong pattern topology, or an aggregation over fan-out-inflated rows.

  \item \textbf{A graph-mined rule library.}
We automatically mine preventive rules to guide query generation.
Specifically, an LLM agent analyzes execution-verified failures from the backbone model across training graphs.
The agent extracts one rule per recurring failure mechanism using a standardized prompt.
To ensure broad applicability, an independent generality validator filters these rules to keep them graph-agnostic.

  \item \textbf{Dataset validation.}
On CypherBench, \cypherTech reaches $\wEAcb\%$ overall execution accuracy (\autoref{sec:results}).
This performance marks a $+\dEAcb\%$ improvement over the same-backbone zero-shot baseline.
% \cypherTech also exceeds zero-shot \texttt{sonnet-4.6} ($\sonnetZsEAcb\%$) by $+\dSonnetCB\%$, even though the latter is a much stronger model.
Applying our framework to stronger backbone yields a comparable performance gain.
Furthermore, \cypherTech outperforms the reported RL-fine-tuned 7B Cypher specialist.
% On the independent Neo4j Text2Cypher benchmark, our method improves \texttt{gpt-5.4-mini} by $+8.97\%$ and \texttt{sonnet-4.6} by $+5.88\%$ over zero-shot floors.

\end{itemize}

%% file: sections/02_background.tex
\section{Background}\label{sec:background}\label{sec:bg-task}

A property graph is a directed, labeled multigraph whose vertices and edges carry type labels and key--value properties. Cypher is one of the query languages in graph databases~\cite{francis2018cypher}. 
We study the schema-grounded formulation. Given a natural-language question $q$ and the schema $S(G)$ of a target graph $G$, which specifies node labels, relationship types, and property keys, we aim to produce, via an LLM, a Cypher query $\hat{c}$ that, when executed on $G$, returns the answer that $q$ asks for.

\noindent \textbf{Execution accuracy.}\label{sec:bg-ea}
Execution accuracy (EA) is the fraction of questions whose predicted query returns a result multiset equal to the gold answer.
It is calculated as
\[ \mathrm{EA}=\frac{1}{n}\sum_{i=1}^{n}\mathbf{1}\!\left[\, \mathrm{eq}\big(\llbracket \hat{c}_i\rrbracket_{G_i},\ \llbracket c^\star_i\rrbracket_{G_i}\big)\,\right], \] where $\llbracket c\rrbracket_G$ is the result of executing $c$ on $G$, $c^\star_i$ is the gold query, and $\mathrm{eq}$ is the result-equality comparator. A runtime error or timeout counts as a miss.
CypherBench~\cite{feng2025cypherbench} also defines two auxiliaries alongside it: \emph{executability}(the executable percentage), the fraction of predictions that run without a runtime error, and the \emph{provenance-subgraph Jaccard similarity} (PSJS), a partial-credit overlap between the subgraphs that the predicted and gold queries match.

This paper mainly targets execution accuracy.
An LLM first generates a draft query $\hat{c}$, which our proposed solution \cypherTech then checks. We call $\hat{c}$ a draft for short. \cypherTech rewrites the draft to fix faults by checking the question's intent, the schema, and the execution result.
A fault-free draft is returned unchanged.
\autoref{sec:method} presents the architecture. \autoref{sec:results} and \autoref{sec:ablations} quantify the accuracy gain contributed by each component.

%% file: sections/03_method.tex
\section{The \cypherTech Layer}
\label{sec:method}

\subsection{Pipeline Overview}\label{sec:method:overview}

\begin{figure*}[t]
\centering
\includegraphics[width=\textwidth]{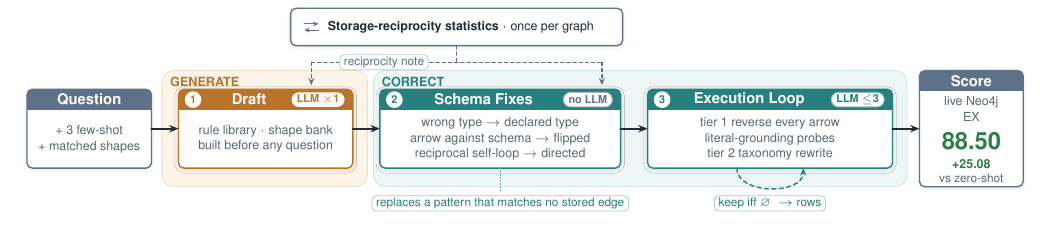}
\caption{The three-stage \cypherTech pipeline. Stage 1 (amber), the Draft, receives the full rule library in its prompt.
Each of Stages 2--3 (teal) rewrites the query when the deterministic test printed on its dashed arc passes.}
\label{fig:pipeline}
\end{figure*}

Our method, namely \cypherTech, is built to improve execution accuracy.
It runs a sequence of checks on the generated Cypher query and update the query when a fault is found.
A straightforward check is to execute the query on a graph database and read any error logs returned by the database. However, only syntax-level errors can be captured in this way. A faulty query may be executed without errors but return empty or wrong results.
To improve the Cypher accuracy, our solution inspects multiple types of faults in three stages shown in \autoref{fig:pipeline} and \autoref{alg:pipeline}: Draft, Schema Fixes, and Execution Loop.

The Draft produces a candidate query (namely draft query for simplicity) at the initial stage.
Its prompt carries the static rule library and two kinds of retrieved exemplars, few-shot and shape (\autoref{sec:method:retrieval}).
The rule library lists the recurring fault patterns so that certain faults can be avoided at the drafting stage.
Schema Fix (\autoref{sec:method:normalize}) resolves faults whose correct form can be determined by the schema, such as an invalid relationship type.
The Execution Loop (\autoref{sec:method:exec}) runs the query and reacts when the query raises an error or returns nothing.
To this end, when the result is empty the loop first tries reversing relationship arrow and keeps the reversal only if it returns data. Otherwise it asks the LLM to rewrite, which replaces the failing query.

\autoref{alg:pipeline} presents the pseudocode of the pipeline.
\textsc{Retrieve} gathers the few-shot and shape exemplars.
\textsc{Draft} then generates the candidate query based on the question, the schema, the rule library, and the exemplars.
\textsc{FixRelTypes}, \textsc{AlignDirections}, and \textsc{RedirectSelfLoops} apply the Schema Fixes in sequence.
\textsc{RepairLoop} then runs the Execution Loop and returns the final query.

\begin{algorithm}[t]
\caption{The \cypherTech Algorithm}
\label{alg:pipeline}
\begin{algorithmic}[1]
\Require question $q$, graph $G$, schema $S(G)$, rule library $R$, train-graph exemplar corpus $D$, reciprocity verdicts $\Sigma$
\State $E \gets \textsc{Retrieve}(q, D)$ \Comment{few-shot + shape exemplars}
\State $c \gets \textsc{Draft}(q, S(G), R, E, \Sigma)$ \Comment{stage 1}
\State $c \gets \textsc{FixRelTypes}(c, S(G))$ \Comment{stage 2}
\State $c \gets \textsc{AlignDirections}(c, S(G))$ 
\State $c \gets \textsc{RedirectSelfLoops}(q, c, \Sigma)$ 
\State \Return $\textsc{RepairLoop}(q, c, G)$ \Comment{stage 3, \autoref{alg:loop}}
\end{algorithmic}
\end{algorithm}

\begin{table}[t]
  \centering
  \footnotesize
  \setlength{\tabcolsep}{2.5pt}
  \caption{Reliability designs for LLM query generation. \new{Schema = declared node labels, relationship types, and edge directions.} Stored = stored-data statistics or content. \new{Exec. = execution feedback.} Commit = model-issued rewrites must pass a deterministic test.}
  \label{tab:positioning}
  \begin{tabular}{@{}lcccc@{}}
    \toprule
    Design & Schema & Stored & Exec. & Commit \\
    \midrule
    Constrained decoding~\cite{scholak2021picard} & \ding{51} & -- & -- & n/a \\
    Exec.-triggered rewriting~\cite{wang2025macsql} & -- & -- & \ding{51} & -- \\
    Judge-guarded correction~\cite{gong2025sqlens} & \ding{51} & \ding{51} & \ding{51} & -- \\
    Content-informed refinement~\cite{ren2025power} & \ding{51} & \ding{51} & \ding{51} & -- \\
    Schema direction corrector~\cite{langchain2026cyphercorrector} & \ding{51} & -- & -- & n/a \\
    Calibrated abstention~\cite{chen2025abstention} & \ding{51} & -- & -- & n/a \\
    Trained specialists~\cite{su2025cypherri,zhang2026cyphersmith} & -- & -- & -- & n/a \\
    \midrule
    \cypherTech & \ding{51} & \ding{51} & \ding{51} & \ding{51} \\
    \bottomrule
  \end{tabular}
\end{table}

\subsection{Transferable Error Taxonomy and Rule Library}\label{sec:method:retrieval}

Recurring errors in Cypher draft generation can be identified and summarized in advance to assist the Draft and Execution Loop. 
This section introduces three artifacts: a taxonomy, a rule library, and a set of exemplars. 
A taxonomy categorizes recurring fault families and guides the Execution Loop's correction prompt.
A mined rule library reformulates these faults as instructions in the Draft prompt to guide query generation.
Retrieved exemplars show the model what a correct query for a similar question looks like.

\noindent \textbf{The taxonomy.}
Taxonomy-driven correction is first established in text-to-SQL~\cite{chaturvedi2025sqlofthought,gong2025sqlens}, but certain Cypher faults cannot be captured by the SQL error taxonomy. For example, a Cypher might use a label or relationship type that the schema never declares. This Cypher can still be parsed and silently returns no data.
Hence, CypherBench~\cite{feng2025cypherbench} annotates ten classes over sampled failures and GQLBench~\cite{su2026gqlbench} adapts that scheme to thirteen classes for NL2GQL. 
While both benchmarks diagnose where models fail, neither uses these failure patterns to correct queries.
We therefore distil observed text-to-Cypher failures into a graph-native taxonomy of eight families. The taxonomy includes: syntax, direction, schema, entity, pattern, aggregation, filter, and return. It spans $32$ fine-grained codes (\autoref{fig:taxonomy}). Those codes steer Cypher rewrite rather than just name the fault.
The tier-2 correction prompt of the Execution Loop lists all $32$ codes and asks the model to name a code, state a root cause, and then write the corrected query (\autoref{sec:method:exec}).
The last column of \autoref{fig:taxonomy} names the mechanism.

\begin{table*}[t]
\centering
\footnotesize
\setlength{\tabcolsep}{6pt}
\renewcommand{\arraystretch}{1.3}
\begin{tabular}{@{}llll@{}}
\toprule
Family & Failure mode & \textcolor{red}{\ding{55}}\,draft\;$\Rightarrow$\;\textcolor{ForestGreen}{\ding{51}}\,fix & Primarily guarded by \\
\midrule
Syntax & unparsable Cypher & {\ttfamily (p:Person\textcolor{red}{]}\;$\Rightarrow$\;(p:Person\textcolor{ForestGreen}{)}} & taxonomy fix \\
Direction & edge traversed backwards & {\shortstack[l]{\ttfamily -[:directedBy]\textcolor{red}{->}\;\\$\Rightarrow$ \ttfamily \textcolor{ForestGreen}{<-}[:directedBy]-}} & self-loop gate; exec reversal \\
Schema & label/type absent from schema & {\ttfamily (:\textcolor{red}{Film})\;$\Rightarrow$\;(:\textcolor{ForestGreen}{Movie})} & Schema Fixes \\
Entity & literal $\neq$ stored value & {\ttfamily '\textcolor{red}{l}incoln'\;$\Rightarrow$\;'\textcolor{ForestGreen}{L}incoln'} & literal grounding \\
Pattern & wrong path shape; a hop is missing & {\ttfamily (m)--(m2)\;$\Rightarrow$\;(m)--\textcolor{ForestGreen}{(:Person)--}(m2)} & rule library \\
Aggregation & misuse inflates rows & {\ttfamily count(p)\;$\Rightarrow$\;count(\textcolor{ForestGreen}{DISTINCT~}p)} & rule library \\
Filter & wrong predicate or operator & {\shortstack[l]{\ttfamily p.mass\_kg\textcolor{red}{>}100\;\\$\Rightarrow$ \ttfamily p.mass\_kg\textcolor{ForestGreen}{>=}100}} & rule library \\
Return & spurious projection /order / limit & {\ttfamily RETURN p\;$\Rightarrow$\;RETURN p\textcolor{ForestGreen}{.name}} & rule library \\
\bottomrule
\end{tabular}
\caption{The eight families of \cypherTech's transferable Cypher error taxonomy.
Each row gives a minimal drafted slip
(\textcolor{red}{\ding{55}}), its committed repair (\textcolor{ForestGreen}{\ding{51}}),
and the guard that primarily catches it, over the running movie and NBA graphs.
The full $32$ fine-grained codes ship with the release artifact.}
\label{fig:taxonomy}
\end{table*}

\noindent \textbf{A mined, dataset-agnostic rule library.}
The taxonomy only helps once a query has already been drafted. The rule library acts earlier. 
A rule is a short instruction that tells the model how to avoid a common query error. These rules are added to the draft prompt to guide the LLM to avoid errors during query generation.
We mine rules from observed failures, rather than hand-write them.
To allow one agent session to cover all failure types, the mining protocol first subsamples the training set.
We group questions by the shape of their MATCH clause, the kind of answer that their RETURN clause produces, and the filter, sort, or aggregate operators that the query uses. We keep at most three questions per group as a reduced set, so that failures from one common combination do not dominate the set.
Within each group, we greedily select three questions by two ranking priorities. 
First, we prioritize questions from training graphs not yet included in the group.
Second, we select the question that covers the most previously unseen relationship types, labels, and properties.

A specific label, relationship type, property, or literal might appears only once across the entire training set. 
\new{For example, names containing quotes or non-ASCII characters. 
If the grouping step discards a query, the pattern miner misses any associated failure modes.}
% Our mining protocol could drop the question that contains that label or relationship type from the reduced set.
The miner then never sees a failure involving that rare label, type, property, or literal.
Thus, the miner cannot summarize a rule for that failure. 
We therefore check which of these elements the reduced set misses, and add back questions until that failue pattern is covered.
The backbone then drafts a query for each question in this reduced set. 
We execute every draft, score each result against the gold answer, and keep every failed draft.
A generated draft query fails in one of three ways: it triggers an execution error, returns empty results while the gold query returns non-empty rows, or yields a different result multiset.
We package each failed draft with diagnostic context: the question, the gold query, the execution outcome, node/relationship labels, and up to three sample result rows from both the draft and the gold query.
This context enables the rule miner to inspect error mechanisms directly without querying the database.

To mine rules, we dispatch a single session of an LLM agent, \texttt{claude-opus-4-8} in our experiments, to process the failure set under a fixed prompt(\autoref{fig:prompt-mine}). 
The prompt directs the agent to read every failure, cluster the failures, write one preventive rule per recurring pattern, and sort the rules into the three tiers of \autoref{tab:rules}.

Each rule identifies the situation in which a mistake arises, explains why the naive query is wrong, and prescribes the corrective pattern. 
An independent validator checks that every rule follows the runtime's rule grammar and clears a generality bar. The validator scans each rule for every node label, relationship type, property, and gold entity name of the graphs. In this way, every surviving rule is written only in generic placeholders.
Mining and validation together produce the rule library. 
One example shows how a failure becomes a rule.
Cypher returns one row per complete match of a pattern.
Matching a central node against multiple one-to-many relationships in a single \texttt{MATCH} clause creates a Cartesian product of its neighbors.
Consequently, each unique pair of neighbors duplicates the central node in the intermediate results.
A \texttt{count(entity)} over those rows then tallies path combinations, not entities, silently inflating an answer. 
\new{The miner reads such failures, finds that they share this one cause, and turns the fix into a rule.}
The mined rule enforces a three-step pattern: identify the target entity, isolate unique instances using an intermediate \texttt{WITH DISTINCT}, and then perform the aggregation.
This ensures the final \texttt{count()} measures unique entities rather than duplicated rows.

The library holds rules in three tiers. Universal rules apply whatever the question asks. Schema-conditioned rules apply when the schema declares the feature. Question-shape rules apply when the question's own wording shows the trigger. \autoref{tab:rules} lists all fourteen rules mined from CypherBench and the fault that each one prevents.
Note that a mined rule is not a syntax tip. It states a decision procedure together with the boundary condition that stops the model from over-applying the procedure, and two more rules show what a decision procedure with a boundary condition looks like.
C6 in Table~\ref{tab:rules} covers questions such as ``which entities share a trait with $X$'', where both hops must meet at the same node. A common draft instead writes them as two patterns, \texttt{(x)--(s1)} and \texttt{(a)--(s2)}. Nothing forces \texttt{s1} and \texttt{s2} to match and the query returns every \texttt{a} with any neighbor of that kind. The rule reuses one variable across both hops, \texttt{(x)--(s)--(a)}, with a boundary condition that keeps $X$ in the answer unless the question says ``other'' or ``besides.''
C2 covers superlatives. A draft fails silently in two ways: it guesses the sort direction fresh on every question, or it applies an unconditional \texttt{LIMIT~1}. A fresh guess is unreliable, since nothing stops the model from picking the opposite direction the next time a question uses the same word. An unconditional \texttt{LIMIT~1} then keeps one of the entities that share this extreme value and drops the rest.
The rule fixes one direction for each superlative word. A question therefore no longer depends on a guess. The rule's boundary condition then checks every entity against the computed extreme.
If the question asks for all tied results, the query return every entity matching that value.

\begin{figure}[t]
\centering
\begin{promptbox}{Rule-mining prompt (condensed)}
% The mining agent receives this brief and nothing else. Its read-only inputs are the failure set, the train-graph schemas, the baseline's accuracy report, and the scoring rule.\par\vspace{5pt}
{\ttfamily\footnotesize\raggedright
TASK. Build an error-prevention rule library for a natural-language-to-Cypher system, by mining the real failures of a bare zero-shot baseline. You work alone, from data. This prompt is your complete specification.\\[3pt]
METHOD. Read every error, not a sample. Cluster by failure mechanism, not by graph or by template id. A mechanism is the property of the prediction that made it wrong and that an instruction could have prevented. Quantify each cluster. Write one rule per mechanism family, merging two mechanisms when a single instruction prevents both, and dropping one-off accidents unless the failure is catastrophic and cheap to prevent. Assign each rule a tier by frequency, harm, and detectability. Use the universal tier for near-universal disciplines whose violation silently corrupts a result, the schema tier when the graph schema determines whether a rule applies, and the question-shape tier when a recognizable surface condition of the question identifies the risk.\\[3pt]
GENERALITY (hard). Never mention a node label, relationship type, or property name of the training graphs, and never quote an entity name from them. Write examples in placeholder identifiers, labels \texttt{A}/\texttt{B}, relationship \texttt{REL}, property \texttt{prop}. Do not mention the benchmark, the graphs, the baseline model, or file names. A rule must read as a timeless instruction about writing Cypher from natural language.\\[3pt]
BUDGET. 12--22 rules in total, of which 4--7 are universal, and at most about 200 words per rule body.\\[3pt]
QUALITY BAR. Optimize for coverage of the observed failures weighted by frequency, for precision, stating each rule's boundary conditions so that the model does not over-apply it, and for prompt economy. Prefer a rule that states a decision procedure over one that states only a prohibition.\par}
\end{promptbox}
\caption{The rule-mining prompt, condensed from the released brief. Tier names follow \autoref{tab:rules}.}
\label{fig:prompt-mine}
\end{figure}

\begin{table*}[t]
\centering
\footnotesize
\color{black}
\setlength{\tabcolsep}{6pt}
\caption{The mined rule library.}
\label{tab:rules}
\begin{tabular}{@{}ll p{0.50\textwidth}@{}}
\toprule
& Rule & The fault it prevents \\
\midrule
\multicolumn{3}{@{}l}{\emph{Universal discipline, in force whatever the question asks}}\\
A1 & Return the columns asked for, and no others & an extra or missing column, most often a sort key that is also returned \\
A2 & Match named things by whole-value equality & substring and paraphrase matching, which pulls in unrelated nodes \\
A3 & Deduplicate the answer entity, then project or aggregate & counts inflated by an entity matching several relationships \\
A4 & Do not filter out null property values & dropped answer rows, and an empty result when no entity has the value \\
A5 & Use only declared elements, in minimal patterns & invented syntax, undeclared labels, and unbound relationship variables \\
\midrule
\multicolumn{3}{@{}l}{\emph{Schema-conditioned, applying when the schema declares the feature}}\\
B1 & Follow the schema's relationship directions & a traversal against the stored direction, which matches nothing \\
B2 & Unwind list-valued properties & nested arrays returned where a flat answer is expected \\
B3 & Read membership from relationship time bounds & comparing against today's date, or reading a null end as never existed \\
\midrule
\multicolumn{3}{@{}l}{\emph{Question-shape, applying when the wording shows the trigger}}\\
C1 & Counting questions & counting matched rows rather than distinct entities \\
C2 & Superlative and top-one questions & a guessed sort direction, and \texttt{LIMIT 1} discarding tied entities \\
C3 & Either/or across two matches & an \texttt{OR} over null tests where a \texttt{UNION} is needed, which miscounts \\
C4 & Comparing two named entities & an extra column leaked by sorting the pair instead of returning a verdict \\
C5 & Count within each group & a per-group count taken at the wrong scope, and zero-count groups vanishing \\
C6 & Shared-neighbor and same-as questions & two node variables that never meet at the shared node \\
\bottomrule
\end{tabular}
\end{table*}

\noindent \textbf{Few-shot and shape exemplars.}
The rule tells what to avoid. Exemplars show what a good query looks like. We supply two types of exemplars, both injected into the Draft prompt alongside the rules.
First, few-shot exemplars carry content. A retriever returns the three train-split questions nearest to the input by embedding similarity~\cite{gao2024dailsql,xie2025opensearchsql}, each paired with its gold Cypher.
The exemplar's schema never matches the test graph's. The exemplar's specific content therefore cannot transfer. However, its clause idioms and pattern shapes can.
Second, shape exemplars carry structure alone. They are a small curated bank of twelve idiomatic Cypher skeletons, such as group-by-count and two-hop chain, written in generic placeholders such as \texttt{NodeLabel} and \texttt{relType}, and selected by keyword overlap with the question.
\autoref{fig:prompt-draft} outlines how the Draft prompt is assembled..

\begin{figure}[t]
\centering
\begin{promptbox}{Draft prompt (stage~1)}
% The Draft call wraps the instruction below with a \textbf{static prefix} (the mined rule library of \autoref{tab:rules} and the storage-reciprocity note of \autoref{sec:method:normalize} on self-loop graphs) and a \textbf{per-question block} (the graph-disjoint few-shot exemplars, the matched shape exemplars, the schema, and the question).\par\vspace{5pt}
{\ttfamily\footnotesize\raggedright
Translate the question to Cypher query based on the schema of a Neo4j knowledge graph.\\[3pt]
$\bullet$~Output the Cypher query in a single line, without any additional output or explanation. Do not wrap the query with any formatting like \textasciigrave\textasciigrave\textasciigrave{}.\\
$\bullet$~Perform graph pattern matching in the MATCH clause where possible.\\
$\bullet$~Avoid listing the same entity twice. If distinct entities share a name, repeat the name as separate entries.\\
$\bullet$~Do not return node objects. Instead, return entity names or properties.\\
\texttt{Graph Schema:}$\langle$schema$\rangle$ \, \texttt{Question:}\,$\langle$question$\rangle$ \,
\texttt{Cypher:}\par}
\vspace{4pt}
\end{promptbox}
\caption{The stage-1 Draft prompt. The quoted instruction is verbatim from the released code, and the labels name the assembled message blocks of \autoref{sec:method:retrieval}.}
\label{fig:prompt-draft}
\end{figure}

\subsection{Schema Fixes and redirect Self-loop}\label{sec:method:normalize}

The rules of \autoref{sec:method:retrieval} steer the Draft. However, a draft query can still use a relationship type that the schema never declares, and the correct form can be derived from the schema.
The Schema Fix rewrite the draft in three steps applied in the order of \autoref{alg:pipeline}. \textsc{FixRelTypes} replaces the undeclared relationship type. \textsc{AlignDirections} reverses the arrow that the schema declares only in the other direction. \textsc{RedirectSelfLoops} makes the undirected self-loop pattern directed when the relationship is stored in both directions.

\textsc{FixRelTypes} repairs the wrong relationship type. The schema declares the full set of types. A type outside this set therefore matches no edge in the graph. The algorithm reads the two endpoint labels written around the undeclared type, and replaces the type with the first one that the schema declares between those two labels.

\begin{exampleO}\label{ex:reltype}
A draft may join \texttt{Movie} to \texttt{Genre} with the type \texttt{inGenre}, which the movie schema never declares. The schema declares type \texttt{hasGenre} between those two labels. The edge is therefore rewritten to it.
\end{exampleO}

\textsc{AlignDirections} repairs the direction of an edge whose two endpoint labels differ. For each relationship type, the schema records the source label and the target label of its edges. When a path traverses a type from \texttt{B} to \texttt{A} but the schema declares it only from \texttt{A} to \texttt{B}, \textsc{AlignDirections} reverses the drafted arrow.

\begin{exampleO}\label{ex:align}
The movie schema declares \texttt{directedBy} only from \texttt{Movie} to \texttt{Person}. A draft that traverses it from a \texttt{Person} node to a \texttt{Movie} node therefore has its arrow rewritten from \texttt{-[:directedBy]->} to \texttt{<-[:directedBy]-}.
\end{exampleO}

\new{
\textsc{RedirectSelfLoops} repairs self-loop relationships.
A self-loop relationship connects nodes of the same label, such as \texttt{hasSpouse} between two \texttt{Person} entities. 
A married couple is often stored as two edges. For example, \texttt{(Alice)-[:hasSpouse]->(Bob)} and \texttt{(Bob)-[:hasSpouse]->(Alice)}. 
An undirected drafted pattern over \texttt{hasSpouse} matches both edges for the same couple. 
A query that counts a person's spouses therefore returns two rows for the one true spouse.
The fault here is a double count. 
Matching such a relationship with an undirected pattern traverses each edge pair twice and doubles the resulting count.
The schema checking module does not distinguishes between source and target labels.
}

\new{
Here, we call a relationship is reciprocal when nearly all of its edges exist bi-directionally in the graph. 
The storage-reciprocity gate rewrites an undirected drafted pattern over a reciprocal relationship into a directed one.
Once per graph at initialization, it measures, for each self-loop relationship, the fraction of edges that are also stored in the reverse direction. A relationship counts as reciprocal once that fraction reaches a certain threshold.
}

\begin{exampleO}\label{ex:reciprocity}
The politics graph stores a diplomatic relation as two edges, one in each direction, and the probe finds a matching reverse edge for $95.2\%$ of that type's edges. An undirected pattern over it reaches each partner twice, doubling a country's partner count, \new{as in \autoref{fig:motivating}.}
\end{exampleO}

A misclassified relationship produces a wrong answer, and since the rewritten query still returns data rather than error messages, the Execution Loop never catches this kind of fault.
To our knowledge, the storage-reciprocity gate is the first text-to-Cypher repair where a measured database statistic can be used to decide a rewrite. Prior systems leave that decision to the model (\autoref{tab:positioning}).

\subsection{The Execution Loop}\label{sec:method:exec}

The Schema Fix acts when the schema can be used to determine the answer. 
The Execution Loop covers the faults that the schema cannot resolve. This module rewrites the draft query when the query returns nothing or raises an error.

\begin{algorithm}[t]
\caption{The Execution Loop}
\label{alg:loop}
\begin{algorithmic}[1]
\Require question $q$, query $c$, graph $G$, budget $L{=}3$
\State $\mathit{seen} \gets \{c\}$; \quad $\mathit{prev} \gets$ none; \quad $\mathit{notes} \gets \varnothing$
\For{$i \gets 1$ \textbf{to} $L$}
  \State $(\mathit{rows}, \mathit{err}) \gets \textsc{Execute}(c, G)$
  \If{$\mathit{rows} \neq \varnothing$ and no $\mathit{err}$} \Return $c$
  \EndIf
  \If{$\mathit{rows} = \varnothing$ and no $\mathit{err}$}
    \State $c_{\mathrm{flip}} \gets \textsc{ReverseAllDirections}(c)$ \Comment{tier 1}
    \If{$\textsc{Execute}(c_{\mathrm{flip}}, G) \neq \varnothing$} \Return $c_{\mathrm{flip}}$
    \EndIf
    \State $\mathit{notes} \gets \textsc{ProbeLiterals}(c, G)$ \Comment{bounded point lookups}
  \EndIf
  \State $(\mathit{code}, c') \gets \textsc{TaxonomyCorrect}(q, c, \mathit{err}, \mathit{notes})$ \Comment{tier 2}
  \If{$c'$ is empty or $c' = c$ or $c' \in \mathit{seen}$} \Return $c$ \Comment{no-change and ping-pong stopping rules}
  \EndIf
  \State add $c'$ to $\mathit{seen}$
  \If{$\mathit{code} \neq \text{none}$ \textbf{and} $\mathit{code} = \mathit{prev}$} \Return $c'$ \Comment{repeated diagnosis, keep newest}
  \EndIf
  \State $\mathit{prev} \gets \mathit{code}$; \quad $c \gets c'$
\EndFor
\State \Return $c$
\end{algorithmic}
\end{algorithm}

\autoref{alg:loop} describes the Execution Loop in three parts.
Lines 1--9 and line 12 are the two-tier loop. 
% \del{They apply Tier~1's relationship direction flip, accept the change only if it produces non-empty results, and fall back to a Tier~2 model call otherwise.}
Line 10 is literal grounding. 
% It probes the database about the query's literals, and appends what it finds to the message that tier 2's call sends to the model.
Lines 13--20 are the stopping rules. 
% They reject an empty, unchanged, or already-tried rewrite, stop the loop once the model repeats a diagnosis, and return the last query when the max attempts are reached.

\noindent \textbf{The two-tier loop.}
A failed draft query enters a loop at most three iterations.
A draft fails when it raises an error or returns no rows.
The first tier reverses arrow in the query at once, re-executes, and keeps the flip only if data appear.
Tier 1 targets the direction faults that the Schema Fixes above did not already resolve, since re-execution is the only check available for those faults.
Tier~1 triggers exclusively on empty results.
Drafts raising execution errors bypass the direction flip and proceed directly to Tier~2, as reversing relationship arrows cannot resolve runtime errors.
The second tier calls the model only when the re-execution yields nothing or the draft raises a runtime error, sending it the correction prompt (\autoref{fig:prompt-correct}).
The model call includes the input question, the failed query, and the execution error.
For empty-result failures, the prompt additionally incorporates the literal-grounding notes described below.
The prompt requires the model to name a code from the error taxonomy (\autoref{tab:taxcodes}), state the cause in one sentence, and then write the corrected query. The named code is kept across iterations and feeds the stopping rules below.

\begin{table}[t]
\centering
\small
\setlength{\tabcolsep}{6pt}
\renewcommand{\arraystretch}{1.2}
\caption{The eight families of the correction taxonomy. The tier-2 prompt asks the model to name a code from these families before it emits a rewrite. Each family heads several of the $32$ fine-grained codes, whose full list ships with the artifact.}
\label{tab:taxcodes}
\begin{tabularx}{\columnwidth}{@{}l l >{\raggedright\arraybackslash}X@{}}
\toprule
Code & Family & What it covers \\
\midrule
\texttt{SYN} & Syntax & Cypher the engine rejects. \\
\texttt{DIR} & Direction & Edge traversed against its stored direction. \\
\texttt{SCH} & Schema & Label, type, or property not in the schema. \\
\texttt{ENT} & Entity & Literal matching no stored value. \\
\texttt{PAT} & Pattern & Traversal shape wrong for the question. \\
\texttt{AGG} & Aggregation & Wrong grouping, count, or \texttt{DISTINCT}. \\
\texttt{FIL} & Filter & Over- or under-restrictive predicate. \\
\texttt{RET} & Return & Wrong projection, ordering, or limit. \\
\bottomrule
\end{tabularx}
\end{table}

\begin{figure}[t]
\centering
\begin{promptbox}{Taxonomy-correction prompt (stage~3, tier~2)}
The call sends two system messages. The \textbf{first} carries the graph schema and, on self-loop graphs, the storage-reciprocity note. The \textbf{second} is an expert-debugger instruction carrying the full $32$-code taxonomy (\autoref{tab:taxcodes}), for example \texttt{[DIR.1]} reversed direction, \texttt{[SCH.2]} wrong relationship type, \texttt{[ENT.2]} case error, \texttt{[AGG.3]} missing \texttt{DISTINCT}, and \texttt{[RET.3]} wrong \texttt{ORDER BY}/\texttt{LIMIT}. It asks the model to name the primary code, state the root cause in one sentence, and emit the corrected query, in the fixed \texttt{ERROR:} / \texttt{CAUSE:} / \texttt{FIX:} format. Family-specific reminders attach to the instruction, such as checking \texttt{[DIR.1]} against the schema's relationship list, using only schema names for \texttt{[SCH.*]}, and verifying types and directions on any empty result.\par\vspace{5pt}
\textbf{User message.} The question, the failed query, and an error signal: the database error message or, for an empty result, a fixed message that asks the model to re-check relationship types, directions, string-literal case, and property names, followed by the literal-grounding notes described below. On the empty-result branch, the literal-grounding notes described below are appended to that error signal.
\end{promptbox}
\caption{The stage-4 tier-2 correction prompt.}
\label{fig:prompt-correct}
\end{figure}

\noindent \textbf{Literal grounding.}
A literal in the query can differ from the stored value, such as a truncated title or a re-cased name does. The query then matches nothing, and an empty result does not show the model how the value is actually spelled.
To solve this problem, \cypherTech grounds query literals against the database. This acts as the Cypher counterpart to text-to-SQL value linking~\cite{talaei2025chess}.

We ground query literals by extracting \emph{literal bindings}. A literal binding is a triple (label, property, value) that states which value the query expects on which property of which node type. For example, \texttt{(:Movie \{name:'Lincoln'\})} yields the binding (\texttt{Movie}, \texttt{name}, \texttt{'Lincoln'}).
The extraction follows two steps:
First, we locate the property and value from inline maps (a property filter written directly inside the node pattern, e.g., \texttt{\{name:'Lincoln'\}} in \texttt{(:Movie \{name:'Lincoln'\})}), equality checks, case-insensitive comparisons, or \texttt{IN} lists.
Second, we resolve the node label. Inline maps supply their own label, while other filters inherit it from the variable's declaration (e.g., reading \texttt{Person} from \texttt{(p:Person)} for \texttt{p.name}).
If a variable has no declared label, we probe all labels in the database.
For each query, we evaluate up to three variable bindings, running at most two verification queries per binding.
The first query tests whether any node matches the candidate value exactly on the given property.
If this check returns no results, we execute a second query for fuzzy candidates and generate a diagnostic note.
This fallback query extracts the longest token from the failed literal and retrieves up to three distinct stored values containing that token as a substring.
If candidate values are found, the note presents them and instructs the model to select an exact match.
Otherwise, the note reports that no similar values exist, directing the model to reconsider the property key or node label.
These diagnostic notes are appended to the repair prompt alongside the failed query.
While text-to-SQL systems often maintain precomputed value indices for entity linking, we leave such offline indexing to future work.

\begin{exampleO}\label{ex:grounding}
A question about films from the United States drafts the filter \texttt{\{name:\,'United States'\}}. Because no country carries that value, the query returns nothing. The second query takes the word \texttt{'United'} and finds the stored \texttt{'United States of America'}. The note reports \texttt{'United States'} as absent, lists the stored \texttt{'United States of America'}, and asks the correction step to copy it exactly.
\end{exampleO}

\noindent \textbf{Loop.}
The rewritten query is re-executed on the next iteration. 
Three stopping rules end the loop early (\autoref{alg:loop}): an empty or unchanged rewrite, a repeated query, or the same error code twice.
The first two return the query that the loop currently holds, because the rewrite that they reject is either absent or already tried.
The third returns the newest rewritten query, because that rewrite is new and only the model's diagnosis has repeated.

\noindent 

\subsection{Cost budget}
The preparation cost is paid once. We mine the rule library and the exemplar corpus from the training split, embed the exemplars for retrieval, and measure the storage-reciprocity statistics on each graph.
Answering a question then costs one embedding lookup, one Draft call, and at most $L{=}3$ repair calls.
The database work per question is also a small constant. Each loop iteration runs at most two executions, and each repair call adds at most six point probes, two for each of at most three literal bindings.

%% file: sections/04_setup.tex
\section{Experimental Setup}\label{sec:setup}

We evaluate \cypherTech on two execution-scored text-to-Cypher datasets, CypherBench~\cite{feng2025cypherbench} and the Neo4j Text2Cypher dataset~\cite{ozsoy2025text2cypher}.

\noindent \textbf{CypherBench.}\label{sec:setup-bench}
The released test set holds $n{=}\nQ$ questions over the seven graphs of \autoref{tab:main}, with train and test graphs disjoint.

\noindent \textbf{Neo4j Text2Cypher.}\label{sec:setup-neo4j}
We audit the released test split before use and remove three kinds of questions.
A gold may fail to execute, may return nothing, or may belong to a set of golds that answer the same question differently. 
We drop the entire set because none of its members can be verified as correct.
We keep $1{,}003$ questions over $11$ databases.

{\emergencystretch=3em
\noindent \textbf{The few-shot corpus and rule library.}\label{sec:setup-protocol}
The few-shot corpus and the rule library both come only from the train split of the dataset under evaluation.
On CypherBench, the corpus holds $8{,}534$ pairs from the four train graphs, and the library is mined from \texttt{gpt-5.4-mini}'s execution failures on those graphs.
Exemplars are retrieved from the corpus by cosine similarity over \texttt{text-embedding-}\allowbreak\texttt{3-small} embeddings (\autoref{sec:method:retrieval}).
The mining protocol is part of the method (\autoref{sec:method:retrieval}). 
We run the same protocol on the Neo4j Text2Cypher training split. This produces $13$ rules and a case base used to retrieve exemplars.\par}

% \noindent \textbf{Scoring.}\label{sec:setup-metric}
% We report EX$_{\mathrm{CB}}$ on both datasets (\autoref{sec:bg-ea}).
% After normalization, result rows compare as a multiset. Columns match up to permutation, and row order counts only when the gold query orders its result.

\noindent \textbf{Backbones and runs.}\label{sec:setup-model}
Every configuration runs on a fixed backbone, \texttt{gpt-5.4-mini} and \texttt{claude-sonnet-4.6}.
The \emph{zero-shot baseline} sends the official CypherBench zero-shot prompt (schema and question only) to the same model.
\cypherTech executes queries on a live Neo4j instance of the graph under test.
Every draft's execution feedback during repair and every reported score therefore come from database execution.

%% file: sections/05_results.tex
\section{Main Results}\label{sec:results}

Our evaluation asks four questions: whether the layer beats the same backbone that it wraps, whether each component contributes, whether the few-shot corpus and the rule library, mined from train graphs alone (\autoref{sec:setup-protocol}), transfer across CypherBench's seven vocabulary-disjoint evaluation graphs, and what the layer costs per question, in model calls and database work (\autoref{sec:abl-cost}).

\subsection{CypherBench}\label{sec:results-cb}

\cypherTech reaches $\wEAcb\%$ overall EX$_{\mathrm{CB}}$ over all $n{=}\nQ$ CypherBench test questions, a gain of $+\dEAcb\%$ over the same-backbone \texttt{gpt-5.4-mini} zero-shot floor ($\zsEAcb\%$) on the identical questions (\autoref{tab:main}).
An inference-only layer beats its own backbone by a wide margin on every graph.
\autoref{sec:ablations} orders the components' contributions.

\begin{table}[t]
\centering
\small
\setlength{\tabcolsep}{5pt}
\caption{Results on the released CypherBench test set ($n{=}\nQ$, seven graphs)}
\label{tab:main}
\begin{tabular}{@{}l r r r r@{}}
\toprule
Test graph & $n$ & \shortstack[r]{\texttt{mini}\\zero-shot} & \textbf{ours} & $\Delta$ \\
\midrule
\texttt{nba}                 & 270 & 64.44 & 95.93 & $+31.49$ \\
\texttt{flight\_accident}    & 189 & 71.96 & 98.41 & $+26.45$ \\
\texttt{fictional\_character}& 385 & 51.95 & 79.48 & $+27.53$ \\
\texttt{company}             & 347 & 65.99 & 87.90 & $+21.91$ \\
\texttt{geography}           & 366 & 63.39 & 91.26 & $+27.87$ \\
\texttt{movie}               & 401 & 69.08 & 89.78 & $+20.70$ \\
\texttt{politics}            & 390 & 61.79 & 84.10 & $+22.31$ \\
\midrule
\textbf{Overall} & \textbf{\nQ} & \textbf{\zsEAcb} & \textbf{\wEAcb} & \textbf{$+\dEAcb$} \\
\midrule
\multicolumn{5}{@{}l}{\emph{Published references on the same test set (as reported)}}\\
\multicolumn{3}{@{}l}{\texttt{gpt-4o-mini}, zero-shot~\cite{feng2025cypherbench}}       & 31.43 & \\
\multicolumn{3}{@{}l}{\texttt{qwen2.5-72b}, zero-shot~\cite{feng2025cypherbench}}       & 41.87 & \\
\multicolumn{3}{@{}l}{\texttt{gpt-4o}, zero-shot~\cite{feng2025cypherbench}}            & 60.18 & \\
\multicolumn{3}{@{}l}{\texttt{claude-3.5-sonnet}, zero-shot~\cite{feng2025cypherbench}} & 61.58 & \\
\multicolumn{3}{@{}l}{Cypher-RI, 7B RL fine-tuned~\cite{su2025cypherri}}                & 69.59 & \\
\multicolumn{3}{@{}l}{CypherSmith, data-aug.\ fine-tuned~\cite{zhang2026cyphersmith}}   & 74.80 & \\
\bottomrule
\end{tabular}
\end{table}

EX$_{\mathrm{CB}}$ are improved on every one of the seven test graphs (\autoref{tab:main}).
The per-graph lift ranges from $+20.70\%$ on \texttt{movie} to $+31.49\%$ on \texttt{nba}. The $+\dEAcb\%$ overall gain reflects consistent improvements across graphs rather than an outlier from one graph.
Two graphs \texttt{flight\_accident} and \texttt{nba} exceed $95\%$, with $98.41\%$ and $95.93\%$ respectively.
\texttt{fictional\_character} is the hardest graph for both the zero-shot baseline and \cypherTech.

\autoref{tab:aux} reports two metrics per graph for the final query that each system returns: the executable rate (the share of queries that run without a runtime error) and PSJS, the Jaccard similarity between the subgraphs matched by the predicted and by the gold query
The zero-shot baseline already executes $\zsExecCB\%$ of its queries successfully. 
Consequently, most of its failures stem from queries that run cleanly but return incorrect results, which aligns with its low PSJS score $\zsPsjsCB$.
\cypherTech raises the executable rate to $\execCB\%$ and PSJS to $\psjsCB$, and both gains hold on every graph. Most of the remaining errors are queries that run and return wrong rows (\autoref{sec:abl-errors}).

\begin{table}[t]
\centering
\small
\setlength{\tabcolsep}{4pt}
\caption{Executable rate and PSJS on CypherBench for \texttt{gpt-5.4-mini}, zero-shot and with \cypherTech.}
\label{tab:aux}
\begin{tabular}{@{}l r rr rr@{}}
\toprule
& & \multicolumn{2}{c}{Executable \%} & \multicolumn{2}{c}{PSJS} \\
\cmidrule(lr){3-4}\cmidrule(lr){5-6}
Test graph & $n$ & zero-shot & ours & zero-shot & ours \\
\midrule
\texttt{nba}                 & 270 & 92.22 & 100.00 & 0.809 & 0.983 \\
\texttt{flight\_accident}    & 189 & 93.12 & 100.00 & 0.814 & 0.993 \\
\texttt{fictional\_character}& 385 & 96.36 & 100.00 & 0.660 & 0.901 \\
\texttt{company}             & 347 & 90.78 & 99.71  & 0.772 & 0.923 \\
\texttt{geography}           & 366 & 95.08 & 99.73  & 0.804 & 0.938 \\
\texttt{movie}               & 401 & 94.26 & 99.75  & 0.802 & 0.935 \\
\texttt{politics}            & 390 & 92.05 & 99.74  & 0.793 & 0.905 \\
\midrule
\textbf{Overall} & \textbf{\nQ} & \textbf{\zsExecCB} & \textbf{\execCB} & \textbf{\zsPsjsCB} & \textbf{\psjsCB} \\
\bottomrule
\end{tabular}
\end{table}

We also evaluate raw \texttt{claude-}\allowbreak\texttt{sonnet-4.6} (without \cypherTech), which outperforms standalone \texttt{gpt-5.4-mini}. This comparison tests whether equipping \texttt{gpt-5.4-mini} with \cypherTech allows it to surpass a stronger backbone.
Zero-shot \texttt{sonnet-4.6} reaches \sonnetZsEAcb\%, which is \dSonnetCB\% below \texttt{gpt-5.4-mini} with \cypherTech. With \cypherTech, \texttt{sonnet-4.6}'s performance rises to \sonnetEAcb\%—a gain of +\dSonnetLift\%.

% Using the same pipeline, we evaluate zero-shot \texttt{claude-sonnet-4.6} to test whether our layer enables a cheaper backbone to surpass this frontier model. Its overall EX$_{\mathrm{CB}}$ is $\sonnetZsEAcb\%$, which is $\dSonnetCB\%$ below \cypherTech despite the stronger backbone. 
% Wrapping the same layer around that sonnet-4.6 backbone reaches $\sonnetEAcb\%$ overall, a $+\dSonnetLift\%$ lift over its own zero-shot floor, close to the $+\dEAcb\%$ lift measured on \texttt{gpt-5.4-mini} (\autoref{tab:sonnet}). The layer's gain generalizes across backbones.

\begin{table}[t]
\centering
\small
\setlength{\tabcolsep}{5pt}
\caption{Comparison on \texttt{claude-sonnet-4.6}.}
\label{tab:sonnet}
\begin{tabular}{@{}l r r r r@{}}
\toprule
Test graph & $n$ & \shortstack[r]{\texttt{sonnet}\\zero-shot} & \textbf{ours} & $\Delta$ \\
\midrule
\texttt{nba}                 & 270 & 73.70 & 98.15 & $+24.45$ \\
\texttt{flight\_accident}    & 189 & 76.19 & 99.47 & $+23.28$ \\
\texttt{fictional\_character}& 385 & 68.83 & 94.29 & $+25.46$ \\
\texttt{company}             & 347 & 73.49 & 94.52 & $+21.03$ \\
\texttt{geography}           & 366 & 68.58 & 95.90 & $+27.32$ \\
\texttt{movie}               & 401 & 73.82 & 95.76 & $+21.94$ \\
\texttt{politics}            & 390 & 70.51 & 88.21 & $+17.70$ \\
\midrule
\textbf{Overall} & \textbf{\nQ} & \textbf{\sonnetZsEAcb} & \textbf{\sonnetEAcb} & \textbf{$+\dSonnetLift$} \\
\bottomrule
\end{tabular}
\end{table}

% For external context, the lower block of \autoref{tab:main} lists published accuracies on the same seven test graphs.
% Among 2024-era zero-shot models the ceiling is $61.58\%$~\cite{feng2025cypherbench}, the RL-fine-tuned 7B specialist Cypher-RI reaches $69.59\%$~\cite{su2025cypherri}, and a data-augmentation fine-tune reaches $74.80\%$~\cite{zhang2026cyphersmith}.
% Read against these anchors, \cypherTech's $\wEAcb\%$ exceeds the strongest zero-shot model by about $\dBest\%$, the RL specialist by about $\dRL\%$, and the strongest of the two trained systems by about $\dCS\%$.
\autoref{tab:main} also lists published results on the CypherBench test graphs. The best zero-shot model in the benchmark paper, \texttt{sonnet-3.5}, reaches $61.58\%$~\cite{feng2025cypherbench}. Cypher-RI, a 7B model fine-tuned with reinforcement learning, reaches $69.59\%$~\cite{su2025cypherri}. CypherSmith, a model fine-tuned on synthesized data, reaches $74.8\%$~\cite{zhang2026cyphersmith}.  \cypherTech with \texttt{gpt-5.4-mini} exceeds these three results by about $\dBest\%$, $\dRL\%$, and $\dCS\%$, respectively.

\subsection{Neo4j Text2Cypher}\label{sec:results-neo4j}

The same layer runs on the audited Neo4j Text2Cypher test set of \autoref{sec:setup-neo4j}, with the few-shot corpus and the rule library both rebuilt from this dataset's own train split (\autoref{sec:setup-protocol}).
The layer lifts \texttt{gpt-5.4-mini} from $47.36\%$ to $56.33\%$ and \texttt{claude-}\allowbreak\texttt{sonnet-4.6} from $56.63\%$ to $62.51\%$ (\autoref{tab:neo4j}).
\cypherTech beats the backbone that it wraps in all four dataset-backbone pairs.

\begin{table}[t]
\centering
\small
\setlength{\tabcolsep}{5pt}
\caption{EX (\%) on the Text2Cypher dataset~\cite{ozsoy2025text2cypher}.}
\label{tab:neo4j}
\begin{tabular}{@{}l r r r r@{}}
\toprule
Backbone & $n$ & zero-shot & \textbf{ours} & $\Delta$ \\
\midrule
\texttt{gpt-5.4-mini}      & $1{,}003$ & $47.36$ & $\mathbf{56.33}$ & $+8.97$ \\
\texttt{claude-sonnet-4.6} & $1{,}003$ & $56.63$ & $\mathbf{62.51}$ & $+5.88$ \\
\bottomrule
\end{tabular}
\end{table}

%% file: sections/06_ablations.tex
\section{Ablations and Analysis}\label{sec:ablations}

The results of \autoref{sec:results} establish that the layer works on both datasets. This section asks whether each component contributes, whether the few-shot corpus and the rule library still help on graphs they were never mined from, and whether the gain is just more compute, then closes on what still fails. Experiments below were run on CypherBench.

Every ablation removes one component and re-runs questions. Each number is therefore the accuracy that component contributes on top of everything else (\autoref{tab:ablation}).

\subsection{Removing a component costs accuracy}\label{sec:abl-matrix}

Each of the five components corresponds to a row. 
\texttt{-exec-loop} skips repairing empty or errored drafts. \texttt{-rules} clears all mined rules without altering other components.
\texttt{-few-shot} sets $k{=}0$, stripping every train-graph exemplar while the rule library stays active, and \texttt{-shape} removes only the shape exemplars.
\texttt{-literal} disables the literal-grounding probes (\autoref{sec:method:exec}). Its probes run inside the Execution Loop, and \texttt{-exec-loop} therefore removes them as well.

\begin{table*}[t]
  \centering
  \footnotesize
  \setlength{\tabcolsep}{4pt}
  \caption{Drop-one ablation on the CypherBench (\texttt{gpt-5.4-mini}). Each row removes one component and reports what it contributes. Positive means the component helps that graph. The gain is over the zero-shot floor.}
  \label{tab:ablation}
  \begin{tabular}{@{}l*{8}{c}@{}}
    \toprule
    Component removed & nba & flight & fiction & company & geo & movie & politics & \textbf{Wtd.} \\
    \midrule
    the whole layer (zero-shot baseline) & $+31.49$ & $+26.45$ & $+27.53$ & $+21.91$ & $+27.87$ & $+20.70$ & $+22.31$ & $\mathbf{+\dEAcb}$ \\
    \midrule
    exec-loop  & $+4.81$  & $+6.35$ & $+20.00$ & $+12.97$ & $+9.84$  & $+10.97$  & $+6.67$ & $\mathbf{+10.78}$ \\
    rule library & $+0.00$ & $+3.17$ & $+3.38$ & $+5.19$ & $+6.83$ & $+5.24$ & $+5.13$ & $\mathbf{+4.39}$ \\
    few-shot exemplars   & $+12.59$ & $+2.65$ & $+2.34$  & $+6.34$  & $+4.37$  & $+10.72$ & $+6.67$ & $\mathbf{+6.60}$ \\
    shape exemplars      & $+0.37$  & $+1.06$ & $+0.78$  & $+1.44$  & $+2.73$  & $+3.74$  & $-1.79$ & $\mathbf{+1.24}$ \\
    -literal$^{\ddagger}$ & $+0.37$ & $+2.65$ & $+4.94$ & $+2.31$ & $+3.83$ & $+2.00$ & $+0.00$ & $\mathbf{+2.34}$ \\
    \bottomrule
  \end{tabular}
\end{table*}

\noindent \textbf{Execution feedback contributes the most.}

Removing any single component lowers EX.
Among them, removing the Execution Loop (\texttt{-exec-loop}) causes the largest drop ($10.78\%$).
The remaining ablations \new{demonstrate consistent positive contributions:} few-shot exemplars (\texttt{-few-shot}, $6.60\%$), the rule library (\texttt{-rules}, $4.39\%$), literal grounding (\texttt{-literal}, $2.34\%$), and shape exemplars (\texttt{-shape}, $1.24\%$).
This is consistent with text-to-SQL self-correction, where execution feedback is a central correction signal~\cite{xu2025tssql,wang2025macsql,talaei2025chess}, or a training reward~\cite{zhai2025excot}.
By removing one component at a time while keeping the rest intact, we isolate individual stage contributions. 
We quantify prevention and repair independently rather than confounding their effects.

Among the two kinds of exemplars, few-shot exemplars matter far more. Removing them costs more than five times as much as removing shape exemplars.
Retrieved gold Cypher supplies complete relational idioms that the rule library lacks, whereas shape exemplars provide only query skeletons, which the rule library might already covers in part.

\subsection{The few-shot corpus and rule library transfer}\label{sec:abl-transfer}\label{sec:abl-pergraph}\label{sec:abl-monotone}

Both the few-shot corpus and the rule library are sourced only from the train graphs (\autoref{sec:setup-bench}). 
Our transfer claim applies only to the few-shot corpus and the rule library.
The evidence bears the transfer claim out in every measurement.
EX$_{\mathrm{CB}}$ improves over the zero-shot floor on all seven graphs (\autoref{tab:main}), and removing the Execution Loop or the few-shot exemplars lowers it on every one of them (\autoref{tab:ablation}).

\noindent \textbf{The Execution Loop and the few-shot exemplars help different graphs.}
The overall average masks substantial per-graph differences between the two components.
The Execution Loop primarily benefits the hardest graphs.
It achieves its largest gain on the lowest-scoring graph, \texttt{fictional\_character} ($+20.00\%$), followed by the third-lowest, \texttt{company} ($+12.97\%$).
Only \texttt{movie} ($+10.97\%$) breaks this trend by outgaining the lower-scoring \texttt{politics} ($+6.67\%$).
In contrast, few-shot exemplars yield the largest gains on \texttt{nba} ($+12.59\%$) and \texttt{movie} ($+10.72\%$).

\noindent \textbf{The rule library transfer across backbones.}
\new{
The rule library is mined from the train graphs. 
It transfers to new graphs.
Removing the rule library lowers EX$_{\mathrm{CB}}$ on six of the seven evaluation graphs.
It transfers to a larger backbone.
We mined the library only from \texttt{gpt-5.4-mini}'s failures. Applied unchanged to \texttt{sonnet-4.6}, it raises accuracy from $\sonnetZsEAcb\%$ to $\sonnetEAcb\%$ on all seven graphs (\autoref{tab:sonnet}).
The library can therefore be mined once with a smaller model and reused with larger ones.
}
% \noindent \textbf{The few-shot corpus and rule library transfer across backbones.}
% Our components generalize across different base models.
% We mined the rule library exclusively from failure cases.
% Applying this exact library to \texttt{claude-sonnet-4.6} improves its accuracy from $\sonnetZsEAcb\%$ to $\sonnetEAcb\%$.
% This gain is consistent across all seven graphs (\autoref{tab:sonnet}).
% Furthermore, cross-backbone transfer does not rely on a low baseline.
% Even though \texttt{claude-sonnet-4.6} starts $8\%$ higher than \texttt{gpt-5.4-mini}, its absolute gain remains within $3\%$ of \texttt{gpt-5.4-mini}'s gain.

\subsection{Is the gain just more compute?}\label{sec:abl-cost}

Pipeline gains are often questioned as merely spending more compute.

\noindent \textbf{The call budget is bounded by design.}
\new{The offline rule mining stage runs only once because mining the rule library requires a single agent session on the baseline's failures, and the resulting rule library can be reused for subsequent nl2cypher tasks. Each question requires one initial Draft call and at most three repair calls.}
In contrast, self-consistency voting and multi-path candidate generation scale the number of calls with a configured sampling budget~\cite{wang2023selfconsistency,pourreza2025chasesql,li2025alphasql}.
% In contrast, unbounded agentic frameworks and self-consistency voting scale inference calls without a ceiling~\cite{wang2023selfconsistency,pourreza2025chasesql,li2025alphasql}.
% Moreover, such methods yield inconsistent returns across different backbones and workflows~\cite{guo2025rethinking}.
% These repair calls execute only when a draft encounters an error or returns an empty result.
% Consequently, total calls per question strictly range from one to four, fixed prior to inference.

% \noindent \textbf{The pipeline provides gains beyond scaling backbone capacity.}
% On CypherBench, \cypherTech on \texttt{gpt-5.4-mini} achieves $\wEAcb\%$.
% This outperforms zero-shot \texttt{claude-sonnet-4.6} by $\dSonnetCB\%$ ($\sonnetZsEAcb\%$, \autoref{tab:sonnet}).
% At most four lightweight calls surpass a single expensive call.
% Despite its higher capacity, the larger zero-shot model lacks access to schema statistics and cannot react to execution feedback.
% On Neo4j Text2Cypher, the performance gap narrows ($56.33$ for pipeline-wrapped \texttt{mini} vs. $62.51$ for zero-shot \texttt{claude-sonnet-4.6}, \autoref{tab:neo4j}).
% However, the best configuration on this benchmark remains our pipeline wrapped around the stronger backbone ($62.51$).

\subsection{Error Analysis at $\wEAcb\%$}\label{sec:abl-errors}\label{sec:abl-stoprule}

On CypherBench, the complete system fails on $270$ out of the \nQ{} questions.
These residual errors are concentrated unevenly across domains.
Over half ($52.2\%$) reside in the two most challenging graphs: \texttt{fictional\_character} ($79$ errors) and \texttt{politics} ($62$ errors).

\noindent \textbf{Non-empty results act as weak verifiers.}
While an empty result indicates a flawed query, a non-empty result does not guarantee correctness.
Our guards trigger only on explicit signals: execution errors, empty outputs, or schema-statistic mismatches.
Therefore, a query that executes successfully and returns plausible but incorrect data bypasses all checks.
% On the three most difficult graphs, $80\%$ to $88\%$ of erroneous queries returned non-empty data, escaping the Execution Loop entirely.
% Although empty-result triggering is standard in text-to-SQL correctors (\autoref{sec:related}), addressing these executing-yet-incorrect queries requires specialized semantic verifiers~\cite{gong2025sqlens}.
Empty-result triggering is common in text-to-SQL correctors~\cite{wang2025macsql,talaei2025chess}, and verifiers designed for executing-yet-incorrect queries exist, for example SQLens~\cite{gong2025sqlens}.

%% file: sections/08_related.tex
\section{Related Work}\label{sec:related}

\cypherTech draws on five threads of related work.

\noindent \textbf{Natural-language interfaces to data systems.}
Querying data in natural language predates large language models, through grammar- and ontology-driven systems~\cite{li2014nalir,saha2016athena} and trainable pipelines~\cite{weir2020dbpal}, with a graph-side lineage in question answering over knowledge graphs~\cite{berant2013freebase,unger2012templateqa}. LLMs moved the field to prompting and made text-to-SQL a standing subject at data-management venues~\cite{li2024codes,xie2025opensearchsql,li2024nl2sql360,katsogiannismeimarakis2023survey}.
Reliability itself has entered that agenda. Adaptive abstention makes text-to-SQL dependable by learning when not to answer, abstaining under a calibrated test and handing those questions back~\cite{chen2025abstention}.
\new{On the graph database side, whose language surface has just standardized~\cite{iso2024gql,deutsch2022gqlpgq}, only partial counterparts exist so far, such as schema-based direction correction~\cite{langchain2026cyphercorrector}.}

\noindent \textbf{Self-correction and execution feedback.}
A second thread repairs a draft after it is written, from the iterate-with-self-feedback and execution-feedback recipes~\cite{madaan2023selfrefine,chen2024selfdebug} to text-to-SQL correctors using retry budgets, semantic error taxonomies, trained selection judges, test-driven loops, and critique- or detection-based edits~\cite{wang2025macsql,chaturvedi2025sqlofthought,pourreza2025chasesql,xu2025tssql,xie2024deasql,wang2025dac,gorti2025mscsql,gong2025sqlens,qu2025share}. A persistent finding is that unconditional refinement can introduce new errors.
Execution-triggered rewriting is a standard SQL corrector design. Ported to Cypher, its empty-result trigger does catch one dominant graph fault, a traversal against a schema-directed edge, because that draft matches nothing. It does not catch the fault our motivating example opens with, a reciprocal relationship matched undirected, whose draft returns rows.

Neo4j's LangChain integration ships a schema-based direction corrector~\cite{langchain2026cyphercorrector}. The loop's blanket reversal instead covers the orientations the schema leaves open and commits only on a non-empty re-execution. Guiding correction with a structured error catalogue is established in text-to-SQL, as a formal taxonomy~\cite{chaturvedi2025sqlofthought} and as a database-signal catalogue~\cite{gong2025sqlens}. 
Closest to our use of database-side signals, REDSQL refines drafted SQL by consulting the content of the stored database itself~\cite{ren2025power}. 
A preventive alternative constrains decoding itself, masking grammar- or schema-invalid tokens during generation~\cite{scholak2021picard} or enforcing query templates mined from recurring historical questions~\cite{jivani2025tecod}. 

\noindent \textbf{Prompting and skill conditioning.}
A third thread reshapes the prompt through staged modules with per-difficulty branching, systematic example selection, adaptive agent coordination, and dynamic few-shot selection~\cite{pourreza2023dinsql,gao2024dailsql,talaei2025chess,xie2025opensearchsql}. A complementary direction conditions on reusable procedural priors, namely MCTS over reasoning trajectories~\cite{li2025alphasql} and skill libraries, human-curated or self-evolved~\cite{zhang2026coevoskills,xu2026agentskills,jiang2026sok},

Unlike static decomposition prompts, heavyweight per-question agents~\cite{cao2025apexsql}, or skills self-generated from task descriptions, \cypherTech conditions on a small, dataset-agnostic rule library mined by an LLM agent from execution-verified failures on held-out train graphs. 
A related industrial thread mines reusable knowledge from an agent's interaction history for one stable enterprise database~\cite{chen2025rubiksql}. Our library is instead mined once from execution failures, applied untuned across seven unseen graphs.
Per-item LLM selection of which knowledge to inject is itself established, from reflection-token gating in retrieval~\cite{asai2024selfrag} to skill-retrieval selectors~\cite{li2026skillflow,su2026sra} and question-conditioned prompting~\cite{xie2024deasql}.

\noindent \textbf{Training the generator.}
A fourth thread trains the generator directly, through execution-based RL rewards or supervised, distillation, and synthetic-data routes~\cite{yao2025arctic,pourreza2025reasoningsql,ma2025sqlr1,zhai2025excot,dai2025reexsql,sheng2025cscsql,papicchio2026think2sql,pourreza2024dtssql,raczkowska2026retrysql,hoang2025distillc,qin2025route}. The route is reaching Cypher, through synthetic fine-tuning data~\cite{zhong2025synthet2c,zhang2026cyphersmith} and an RL-trained 7B specialist~\cite{su2025cypherri}, but it presupposes supervision that Cypher offers mainly inside benchmark train splits, and the strongest of these trained systems still trails our untrained layer as reported (\autoref{tab:main}). \cypherTech is fully inference-only and treats the generator as fixed.

\noindent \textbf{The text-to-Cypher landscape.}
Cypher is a widely adopted property-graph language~\cite{francis2018cypher}, and graph-backed retrieval over knowledge graphs has proven effective for global sensemaking~\cite{edge2024graphrag}. CypherBench shows even strong proprietary models plateau below saturation while small models collapse~\cite{feng2025cypherbench}. The Text2Cypher baseline runs a direct generate-and-run pipeline, optionally with schema filtering~\cite{ozsoy2025text2cypher,ozsoy2025schemafiltering}. R3-NL2GQL coordinates fine-tuned sub-models with knowledge-graph alignment on nGQL~\cite{zhou2024r3nl2gql}. SpCQL and Mind-the-Query build Cypher datasets~\cite{guo2022spcql,chauhan2025mindthequery}, and SyntheT2C is a synthetic-data framework that ships a smaller medical set~\cite{zhong2025synthet2c}. Text2GQL-Bench adds cross-dialect evaluation~\cite{lyu2026text2gqlbench}. 
Prior Cypher and GQL work reports error categories without acting on them~\cite{feng2025cypherbench,su2026gqlbench}, R3-NL2GQL stages generation across fine-tuned sub-models and a large-model refiner~\cite{zhou2024r3nl2gql}, and the shipped direction corrector lacks an execution-commit gate~\cite{langchain2026cyphercorrector}.

%% file: sections/09_conclusion.tex
\section{Conclusion}\label{sec:conclusion}

\new{We presented \cypherTech, a layer that sits between an LLM and a graph database and checks the Cypher queries. The layer first fixes the errors that can be detected from the schema alone. It then runs every query and asks the LLM to rewrite only when the query fails or returns nothing. 
A repair is accepted only when database verification yields a valid, non-empty result.
Both the rule library for query generation and the error taxonomy for repair are derived from the execution failures on training graphs.}
\new{Across CypherBench test questions, the layer improves \texttt{gpt-5.4-mini} execution accuracy from $\zsEAcb\%$ to $\wEAcb\%$ and \texttt{sonnet-4.6} from $\sonnetZsEAcb\%$ to $\sonnetEAcb\%$.
Both backbones achieve consistent gains on Neo4j Text2Cypher dataset as well.
}

Our framework currently targets Neo4j Cypher. Remaining errors primarily involve queries that execute without runtime failure but return incorrect rows. Future work includes semantic result validation (beyond non-empty checks) and extending our approach to ISO-GQL.